# Macroscopic Hadronic Little Black Hole Interactions

Mario Rabinowitz
Armor Research; lrainbow@stanford.edu
715 Lakemead Way, Redwood City, CA 94062-3922     AR/3

**Abstract**
Although the extraordinary weakness of gravity makes it by far the weakest of the interactions, viewing little black holes (LBH) as a class of elementary particles puts them in a league with hadrons as strongly interacting particles. They interact strongly both in the subatomic and macroscopic realms. An enormous universal gravitational attractive force $\sim 10^{43}/B^2$ Newtons acts between identical black holes for any mass $M \gg M_{Planck}$ at a center-to-center separation of $2 BR_H$ ($B > 1$). For $M \sim M_{Planck}$, there is a comparably large repulsive force. The Hawking model of LBH radiation permits a luminosity from a single LBH comparable to that of the entire universe, whereas this luminosity is greatly attenuated in the Rabinowitz model. In interacting with each other, and with large macroscopic bodies such as stars, neutron stars, and planets, LBH can exhibit strong interactions with large-scale manifestations. Although previously dismissed, an LBH is a potential candidate in accounting for the 1908 devastation of Tungus Siberia, since important LBH interactions were overlooked. LBH passing through neutron star pulsars are capable of causing a sudden change in frequency which may not be fully accounted for by other theories. The existence of black holes is also discussed.

## 1. INTRODUCTION

The weakness of gravity is illustrated by the ratio of the gravitational force to the electric force of $2.4 \times 10^{-43}$ between two electrons (mass $9.1 \times 10^{-31}$ kg) and $8.0 \times 10^{-37}$ between two protons (mass $1.7 \times 10^{-27}$ kg). However two little black holes (LBH) each the size of a nucleon ($R_H \sim 10^{-15}$ m) have as much mass as a mountain ($10^{12}$ kg $\approx 10^9$ ton), completely turning this ratio around to $\sim 10^{41}$ which is well beyond normal strong interactions. The word "little" as used herein refers to the black hole radius, rather than its mass. As we shall see in Sec. 3.2, for very low mass LBH, the repulsive radiation force cannot be neglected.

The question of whether or not black holes exist is discussed in Sec. 4. If black holes exist, LBH are primarily, if not exclusively made in the milieu of the high energies and high pressures of the big bang. This can be understood by looking at the extremely high density of LBH. To create a black hole (BH), an object of mass M must be crushed to a density

$$\rho = \frac{M}{\frac{4\pi}{3}R_H^3} = 7.3 \times 10^{79} M_{kg}^{-2} \text{ kg/m}^3, \text{ where} \quad (1.1)$$

$$R_H = 2GM/c^2 = 1.48 \times 10^{-27} M_{kg} \text{ m} \quad (1.2)$$

is the Schwarzchild radius, often also called the horizon of the BH. Thus a $10^{-3}$ kg LBH has $R_H \sim 10^{-30}$ m and $\rho \sim 10^{86}$ kg/m$^3$ ($10^{83}$ g/cm$^3$). A LBH the size of a nucleon ($R_H \sim 10^{-15}$ m) has a mass of $10^{12}$ kg and density $\rho \sim 10^{56}$ kg/m$^3$ ($10^{53}$ g/cm$^3$).

LBH can be characterized by a few variables such as mass, angular momentum, and electric charge just as is done with ordinary elementary particles. Nathan Rosen (1989 a, b) [of the Einstein, Podolsky, Rosen paradox] was one of the first scientists that considered a possible connection between elementary particles and LBH. LBH of Planck mass (2.2 x $10^{-8}$ kg and $10^{-33}$ m) with charges $\pm\frac{1}{3}e, \pm\frac{2}{3}e, \text{and} \pm e$ were the starting point of his investigation. This paper shall only consider uncharged LBH.

## 2. LITTLE BLACK HOLE RADIATION MODELS

Radiation may be emitted from black holes in a process differing from that of Hawking radiation, $P_{SH}$, which has been undetected for over 25 years. As derived in the Rabinowitz tunneling model (1999 a, b, c), beamed exhaust radiation $P_R$ tunnels out from a LBH due to the field of a second body, which lowers the LBH gravitational potential energy barrier and gives the barrier a finite width. Particles can escape by tunneling (as in field emission) or over the top of the lowered barrier (as in Schottky emission). The former is similar to electric field emission of electrons from a metal by the application of an external field.

Although $P_R$ is of a different physical origin and has different physical consequences than Hawking radiation, it is analytically of the same form. The tunneling power radiated from a LBH is:

$$P_R \approx \left[\frac{\hbar c^3}{4\pi GM}\right]\frac{\langle\Gamma\rangle c^3}{4GM} = \left[\frac{\hbar c^6 \langle\Gamma\rangle}{16\pi G^2}\right]\frac{1}{M^2} \sim \frac{\langle\Gamma\rangle}{M^2}\left[3.42 \times 10^{35} \text{ W}\right] = 60\Gamma P_{SH}$$
(2.1)

where M in kg is the mass of the LBH, $\Gamma$ is the transmission probability $\approx$ WKBJ tunneling probability $e^{-2\Delta\gamma}$ (Rabinowitz, 1999 a, b, c). One might be disposed to challenge the use of the WKBJ approximation in solving the tunneling problem since most of the radiation wavelengths are $\sim R_H$ as measured at large distances. The emitted radiation does not undergo a gravitational red shift in tunneling through the barrier. However after tunneling the barrier,

$$\nu_{observer} = \nu(r_{emission\,site})\left[1 - \frac{2GM/c^2}{r}\right]^{1/2} = \nu(r_{emiss\,site})\left[1 - \frac{R_H}{r}\right]^{1/2}, \quad (2.2)$$

where $\nu_{observer}$ is the frequency detected by the observer at a very large distance from the BH, and $\nu(r_{emission\,site})$ is the frequency at the radial distance r from the center of the BH just outside the barrier. Therefore, the wavelengths for barriers near $R_H$ are $<< R_H$, and the semi-classical WKBJ approach is justified.

For an isolated LBH with M ≳ $10 M_{Planck}$ = 2.2 x $10^{-7}$ kg, the Hawking model predicts

$$P_{SH} = \left[\frac{\hbar c^6}{960\pi G^2}\right]\frac{1}{M^2} \sim 10^{47}\,W, \quad (2.3)$$

with a power density of $\frac{P_{SH}}{4\pi R_H^2} \sim 10^{105}\,W/m^2 = 10^{101}\,W/cm^2$. The Hawking high frequency luminosity of such an LBH is comparable to the visible luminosity of the entire universe:

$P_{univ}$ ~ (~$10^{26}$ W/star)(~$10^{12}$ stars/galaxy)(~$10^9$ to $10^{12}$ galaxies)

~$10^{47}$ to $10^{50}$ W. (2.4)

The argument cannot be made that there are presently no LBH with such a small mass. Even though in the Hawking model all LBH created in the big bang with M ≤ $10^{12}$ kg would have evaporated by now, large numbers of originally more massive LBH can now have evaporated down to ≳ $M_{Planck}$. Such a glaringly large luminosity is not expected from $P_R$. A very close encounter of two LBH would be required, and as we shall see, this is highly unlikely due to the beamed radiation between them that produces a repulsive force.

Another argument that favors $P_R$ is that the radiation is due to a tunneling process and not an information-voiding Planckian black body radiation distribution. Thus $P_R$ can carry information related to the formation of a BH, and avoid the information paradox associated with Hawking radiation. Also the reduced radiation of $P_R$ allows LBH to be candidates for the dark matter, i.e. 95% of the missing mass of the universe. For Hawking that many LBH would fry the universe. He concludes that his LBH can't be more than one-millionth of the mass of the universe. Belinski (1995), a noted authority in the field of general relativity, unequivocally concludes "the effect [Hawking radiation] does not exist." Many profound consequences can result from a change in the model of BH radiation.

### 3. FORCES BETWEEN NEUTRAL BLACK HOLES
**3.1 Universal Attractive Force**

The attractive force between two identical black holes (BH) of mass $M \gg M_{Planck}$ at a separation of $2BR_H$ (B > 1) is

$$F_A \approx \frac{GM^2}{r^2} = \frac{GM^2}{[2BR_H]^2} = \frac{GM^2}{\left[2B\left(\frac{2GM}{c^2}\right)\right]^2} = \frac{c^4}{16B^2G} \sim \frac{10^{43} \text{N}}{B^2}. \quad (3.1)$$

This is a universal attractive force that acts between two identical black holes of any mass at all separations $\gg R_H$, provided that their separation is scaled in terms of the same multiple of $R_H$.

The expression (3.1) is only approximate since the close proximity of two BH distorts the horizons on the adjoining sides of the BH, suppressing $R_H$. In addition, the Einsteinian effective potential of a BH is ~ four times stronger than the Newtonian potential near the BH, although the two are approximately equal for B > 10, i.e. for r > 10 $R_H$ (Rabinowitz, 1999c). Furthermore, eq. (3.1) neglects the radiative repulsive force due to the tunneling radiation between BH, which is discussed next.

**3.2 Repulsive Radiative Force**

The repulsive radiative force between two black holes is

$$F_R \sim -c\left[\frac{dM}{dt}\right] \approx -c\left[\frac{-P_R}{c^2}\right] = \frac{P_R}{c} \approx \frac{1}{c}\left[\frac{\hbar c^6 \langle \Gamma \rangle}{16\pi G^2}\right]\frac{1}{M^2}, \quad (3.2)$$

where $P_R$ is given by eq. (2.1), and the transmission coefficient $\Gamma \approx$ the tunneling probability $e^{-2\Delta\gamma}$ for LBH (Rabinowitz, 1999 a) since the emitted particle velocity $\approx c$ on both sides of the barrier. In the 0 angular momentum case with the origin at the center of mass of masses M and $M_2$:

$$\Delta\gamma = [\vec{b}_2 - \vec{b}_1]\left\{\frac{2\mu}{\hbar^2}\left[Gm\left(\frac{M}{r} + \frac{M_2}{r_2}\right) - E\right]\right\}^{\frac{1}{2}}, \quad (3.3)$$

where $b_2$ and $b_1$ are the turning points of the potential barrier, the reduced mass $\mu = \dfrac{MM_2}{M+M_2}$, and the total energy $E = \dfrac{-GmM}{r-b_1} + \dfrac{-GMM_2}{r_2+b_1}$. For $M = M_2$, and primarily on-axis tunneling, eq. (3.3) reduces to

$$\Delta\gamma = [2b]\left\{\dfrac{M}{h^2}\left[Gm\left(\dfrac{2M}{r}\right) - E\right]\right\}^{\frac{1}{2}}. \qquad (3.4)$$

The average tunneling mass is related to the BH mass through the BH temperature: $\langle m \rangle \approx kT/c^2 \propto 1/M$ (Rabinowitz, 1999 a). So it is not necessary to know in detail the nature of the emitted constituents.

Let us find the LBH mass for which the repulsive and attractive forces are comparable. As given by eqs. (3.1) and (3.2) $F_R \sim F_A$ yields

$$\dfrac{c^4}{16B^2G} \sim \dfrac{1}{c}\left[\dfrac{hc^6\langle\Gamma\rangle}{16\pi G^2}\right]\dfrac{1}{M^2} \Rightarrow M \sim \left[\dfrac{hcB^2\langle\Gamma\rangle}{\pi G}\right]^{\frac{1}{2}}, \qquad (3.5)$$

For $B^2\langle\Gamma\rangle \to \sim 1$, eq. (3.5) yields $M \to \sim M_{Planck}$, since $\langle\Gamma\rangle \xrightarrow[b\to\sim 0]{} \sim 1$ by eq. (3.4).

In the very low probability configurational limit of $\langle\Gamma\rangle \sim 1$, the form of $P_R$ looks like the Hawking radiated power $P_{SH}$, with an important distinction. $P_{SH}$ is omnidirectional and does not yield repulsion in the standard Hawking model, whereas $P_R$ is beamed between the two bodies resulting in repulsion. Two LBH must get quite close for maximum tunneling radiation. In this low probability limit, there is a similarity between the tunneling model and what may be expected from the Hawking model (1974,1975), in that the tidal forces of two LBH would add together to give more radiation at their interface in Hawking's model. This should also produce a repulsive force, though somewhat smaller than from $P_R$, since there is also radiation in all directions.

One should use quantum gravity for such calculations, but it hasn't yet been formulated. As previously discussed, there may be concern regarding the use of semi-classical physics at the Planck scale of $\sim 10^{-35}$ m with energy $\sim 10^{19}$ GeV. However according to eq. (2.2), as measured at large distances, the gravitational red shift substantially reduces the impact of high energies near LBH.

### 4. DO BLACK HOLES EXIST, AND IF SO WHERE?

Though black holes were long considered to be a fiction, most astronomers and astrophysicists believe that their existence is now firmly established. In our own galaxy and in the galaxy NGC 4258, the central dark mass is clearly a verly compact massive object which is most likely a BH. In the case of our galaxy, recent measurements of the velocities of stars as close as 5 light-days from the dynamical center imply a BH of $2.6 \times 10^6$ solar masses, as reported by Genzel (1998).

On an astronomical scale, black holes are thought to be the centers of attraction of galaxies that generate the vast power emitted by quasars, the most luminous objects known in the universe. Quasars are an extreme form of active galactic nuclei, powered by the accretion of matter into supermassive black holes of $10^6$ - $10^{10}$ solar masses. Once these black holes are spun up by infalling matter, up to 43% of the rest mass energy can be tapped if the BH rotation is magnetically coupled to the surrounding interstellar medium. The quasar luminosity (power output) far exceeds the luminosity of its entire galaxy (Davies, 1992).

On the other side of the question, a variation of Einstein's general relativity (EGR) has no black holes, and argues that recent discoveries of quasars with no surrounding material is a data point against EGR. In EGR all fields, except the gravitational field, produce space-time curvature, since the gravitational field is but a consequence of the curvature of space-time. Einstein reasoned that since curvature produced gravity, curvature cannot change gravity, i.e. make more or less gravity. To Einstein this would be double counting. This prohibition in EGR ultimately leads to black holes, in which singularities are the most egregious difficulties.

Hüseyin Yilmaz (1958) made an interesting variation of EGR which avoids not only the singularities but the black holes altogether (Mizobuchi, 1985). Another point in support of Yilmaz is that in particle physics, energy is ascribed to the gravitational field. He assumed that gravitation field energy also produces curvature of space-time by adding a "gravitational stress-energy tensor" to Einstein's equations. Since this term is relatively small in the three major tests of EGR, Yilmaz' General Relativity (YGR) makes essentially the same predictions as EGR for the advance of the perihelion of Mercury, the gravitational red shift, and the bending of starlight (the least accurately measured of the three tests). It appears that EGR and YGR cannot both satisfy the equivalence principle of GR -- at least not in its strong form.

It may not be obvious why including gravitational field energy as a source of space curvature eliminates black holes. An intuitive way to understand this is that the static field and the near field (induction field) of a time-varying gravitational field have negative energy. The radiation field has positive energy. Negative energy gives negative curvature, tending to cancel the positive curvature due to mass. Instead of black holes, YGR has grey holes where the emitted light is greatly red-shifted.

Both EGR and YGR allow for highly compressed objects -- the former with an event horizon (Schwarzchild radius, $R_H$), and the latter without. Much of the analysis in this paper does not depend on the existence of an event horizon, and thus may be valid in either case. The designation LBH herein will include the possibility that little holes may be grey as in YGR, rather than black as in EGR.

Many geophysical and astrophysical processes are not yet well understood. There would be profound implications if it could be established that LBH are the dark matter of the universe, and on rare occasion initiate tremors and trigger quakes. A testable LBH model of sporadic tremors and quakes is explored to determine under what conditions LBH may be relevant to geophysical and astrophysical processes.

In the numerical examples which follow, $M \sim 10^{12}$ kg is used for illustrating the passage of a LBH through the earth, sun, and neutron stars. For levitating in the atmosphere, $M \sim 10^{-3}$ kg, and as M decreases the LBH is repelled away from the earth before producing destructive radiation. In my model the LBH radiation is a function of both the distance and mass of the second body from the LBH as well as the mass of the LBH, and is greatly attenuated relative to Hawking's. For Hawking it depends only on the LBH mass and would be $5.70 \times 10^9$ W for $M = 10^{12}$ kg. For $M = 10^{-3}$ kg, his would be $5.70 \times 10^{39}$ W, which is so exceedingly high that it would cause devastation. As will be shown next in Sec. 5, the incidence rate of LBH and BL seems to be well-matched, so that ball lightning may be an indicator of the distribution of low mass LBH in the universe.

## 5. LITTLE BLACK HOLE FLUX
### 5.1 LBH Flux on the Earth

For LBH coming to the earth from an extremely large distance in essentially free fall from the edge of the universe ($R_U \sim 1.4 \times 10^{26}$ m), by the conservation of energy we can calculate $v_{bh}$ their velocity at the earth

$$v_{bh} = \left[ v_{LBH}^2 + \frac{2GM_e}{R_e} - \frac{2GM_e}{R_U} \right]^{1/2} \approx \left[ v_{LBH}^2 + \frac{2GM_e}{R_e} \right]^{1/2}, \quad (5.1)$$

where $M_e = 6.0 \times 10^{26}$ kg and $R_e = 6.4 \times 10^6$ m are the earth's mass and radius. If the initial LBH velocity $v_{LBH} = 0$, $v_{bh} \sim 10^4$ m/sec. This is the rationale and LBH velocity used by others in the past.

However, a substantially larger velocity should be used. Since the LBH were created during the big bang, at a large distance from the earth they should be in the cosmic rest frame. The velocity of our local group of galaxies with respect to the microwave background, i.e. with respect to the cosmic rest frame (Turner and Tyson, 1999) is a reasonable velocity $v_{LBH} \sim 6.2 \times 10^5$ m/sec for the LBH with respect to the earth at $R_U$. Thus $v_{bh} \approx v_{LBH}$. It is an interesting coincidence that the free fall velocity from rest at $R_U$ to the sun is $6.2 \times 10^5$ m/sec which is the same as $v_{LBH}$, where $M_{sun} = 2.0 \times 10^{30}$ kg and $R_{sun} = 7.0 \times 10^8$ m. Thus for the sun $v_{bh} \approx \sqrt{2}(6.2 \times 10^5 \text{ m/sec}) = 8.8 \times 10^5$ m/sec. For a neutron star with $M_n = M_{sun} = 2.0 \times 10^{30}$ kg, and $R_n = 10^4$ m, $v_{bh} \approx 1.6 \times 10^8$ m/sec $\approx 0.5$ c, which is close to where relativistic effects become important. This should not be surprising as a neutron star is close to being a BH, where $v_{bh}$ would be c, the speed of light.

The continuity equation for mass flow of LBH when there is a creation rate $S_c$ and a decay rate $S_d$ of mass per unit volume per unit time t is

$$\nabla \bullet (\rho \vec{v}) + \partial \rho / \partial t = S_c - S_d, \quad (5.2)$$

where $\rho$ is the LBH mass density at a given point in the universe, $\vec{v}$ is the LBH velocity, and $\rho \vec{v}$ is the LBH flux density. In steady state, $\partial \rho / \partial t = 0$ since then $\rho$ has no time dependence, though $\rho$ may have a spatial dependence. Eq. (5.2) can be converted to an area integral of the flux = a volume integral of the net mass creation-decay rate:

$$\int (\rho \vec{v}) \bullet d\vec{A} = \int (S_c - S_d) dV_t. \quad (5.3)$$

Equation (5.3) integrates to

$$-\rho_{LBH} v_{LBH} A_{far} + \rho_{BL} v_{BL} A_E = (S_c - S_d) V_t, \quad (5.4)$$

where $\rho_{LBH}$ is the mass density of LBH at a distance far from the earth, typical of the average mass density of LBH throughout the universe. $A_{far}$ is the cross-sectional area of a curvilinear flux tube (cylinder) of LBH far from the earth, $A_E$ is the cross-sectional area of the tube where it ends at the earth, and $V_t$ is the volume of the curvilinear flux tube. The symbols $\rho_{BL}$ and $v_{BL}$ represent the number density and velocity of LBH at the earth. The mnemonic subscript BL(ball lightning) is used because in other work (Rabinowitz, 1999 a) a possible connection has been established that LBH can manifest themselves as BL when they go through the atmosphere. This association does not have to be accepted in critically following the derivation presented here. The reader may then just regard references to BL as relating to LBH in the earth's atmosphere.

Because $v_{LBH}$ is high and LBH radiate little until they are near other masses, $S_c$ can be neglected with negligible decay of large black holes to LBH in the volume $V_t$. Similarly, $S_d$ may be expected to be small until LBH are in the vicinity of the earth where most of their evaporation, before they are repelled away by radiative repulsion, is in a volume of the atmosphere $\sim A_E h$, where $A_E$ is the cross-sectional area of the earth, and h is a characteristic height above the earth. At this point it is helpful to convert to number density $\rho_L$ and $\rho_B$, of LBH and ball lightning respectively. The number density decay rate is $\rho_B A_E h/\tau$, where $\tau < \sim$ year is the dwell-time of LBH near the earth. Thus

$$\rho_B = \rho_L \left[ \frac{v_{LBH}}{v_{BL} + (h/\tau)} \right] \frac{A_{far}}{A_E}, \qquad (5.5)$$

which implies that the ball lightning flux is

$$\rho_B v_{BL} = \rho_L v_{LBH} \left[ \frac{v_{BL}}{v_{BL} + (h/\tau)} \right] \frac{A_{far}}{A_E} \approx \rho_L v_{LBH} \left( \frac{A_{far}}{A_E} \right). \qquad (5.6)$$

At large velocities, LBH that do not slow down appreciably due to their large mass or angle of approach, either do not produce sufficient ionization to be seen or do not spend sufficient time in the atmosphere to be observed. These more massive LBH go through the earth and this interaction is covered in Sec. 5.3. As will be shown, the LBH incidence rate matches the estimated BL rate well, but may be too low to account for much quake activity unless the heavier LBH are in re-entrant orbits.

**5.2. Incidence Rate of LBH in the Atmosphere**

In the Rabinowitz model (1999 a, b, c), those LBH that reach the earth's atmosphere and are small enough to have sufficient radiation reaction force to slow them down to the range of $10^{-2}$ to $10^2$ m/sec with a typical value $v_{BL} \sim 1$ m/sec, can manifest themselves as BL. In most cases $h/\tau << v_{BL}$. So Eq. (5.6) implies that the ball lightning current in the atmosphere $\approx$ the LBH current far away. We can thus give a range for the BL flux density

$$\rho_L v_{LBH} < \rho_B v_{BL} < \rho_L v_{LBH} \left( \frac{A_{far}}{A_E} \right). \qquad (5.7)$$

The distribution of LBH masses is not known. Assuming that LBH comprise all of the dark matter, i. e. 95 % of the mass of the universe (Rabinowitz, 1999 a) with 10% of the LBH average mass $\overline{M}_{LBH} \sim 10^{-3}$ kg which can linger in the atmosphere:

$$\rho_L \sim \frac{0.1(0.95 M_{univ} / \overline{M}_{LBH})}{V_{univ}}. \quad (5.8)$$

For $M_{univ} \sim 10^{53}$ kg and $V_{univ} \sim 10^{79}$ m$^3$ (radius of 15 x10$^9$ light-year = 1.4 x 10$^{26}$ m), $\rho_L \sim 10^{55}$ LBH/ $10^{79}$ m$^3$ = $10^{-24}$ LBH/m$^3$. Thus my model predicts that the incidence rate of BL is roughly in the range
$10^{-12}$ km$^{-2}$ sec$^{-1}$ to >~ $10^{-8}$ km$^{-2}$ sec$^{-1}$ for $A_{far}/A_E$ > ~$10^4$. Barry and Singer (1988) give a value of 3 x $10^{-11}$ km$^{-2}$ sec$^{-1}$, whereas Smirnov (1993) estimates 6.4 x $10^{-8}$ km$^{-2}$ sec$^{-1}$ to $10^{-6}$ km$^{-2}$ sec$^{-1}$. In any case the BL incidence is much less than that of ordinary lightning. Even if the percentage were 100%, $10^{-11}$ km$^{-2}$ sec$^{-1}$ is well below the noise level of existing devices such as at large facilities for neutrino detection.

**5.3. Incidence Rate of LBH Through the Earth**

Assuming that 10% of the more massive LBH have an average mass $\overline{M}_{LBH} \sim 10^{12}$ kg, there are ~ $10^{40}$ LBH/ $10^{79}$ m$^3$ = $10^{-39}$ LBH/m$^3$, and with ($A_{far}/A_E$) ~ $10^6$ eqs. (5.7) and (5.8) imply that the flux of these LBH through the earth is ~ $10^{-27}$/m$^2$-sec. Such LBH are too massive to produce enough exhaust radiation to linger in the atmosphere, and so go right through the earth. The earth's diameter is 1.3 x 10$^7$ m which implies that the incidence rate of $10^{12}$ kg LBH is ~ ($10^{-13}$/sec) ~ 1 LBH /$10^5$ year. This can be greatly augmented if the heavier LBH are in re-entrant orbits.

**5.4. Incidence Rate of LBH Through the Sun**

With an average mass $\overline{M}_{LBH} \sim 10^{12}$ kg for 10% of the more massive LBH, gravitational focussing may increase their flux ~ $10^5$ than through the earth to ~ $10^{-22}$/m$^2$-sec. The sun's diameter is 1.4 x 10$^9$ m, implying an incidence rate of ~ $10^{-9}$/sec ~ 1 LBH /10 year through the sun, neglecting LBH re-entrant orbits.

**5.5. Incidence Rate of LBH Through Neutron Stars**

With a gravitational enhancement ~ $10^{10}$ with respect to the earth, the flux is ~ $10^{-17}$/m$^2$-sec for the heavier LBH through neutron stars. The typical diameter of neutron stars of ~ $10^4$ m, implies an incidence rate of ~ $10^{-9}$/sec ~ 1 LBH /10 year through neutron stars, neglecting re-entrant orbits for the LBH.

# 6. INTERACTION OF A LITTLE BLACK HOLE AS IT PASSES THROUGH MATTER

The change in momentum of a particle of mass m of negligible initial velocity due to the impulse imparted by a LBH of mass M and velocity $v_{bh}$ as it passes by is

$$m\Delta v = (F)\Delta t = \left(\frac{GMm}{b^2}\right)\frac{2b}{v_{bh}} = \frac{2GMm}{bv_{bh}} \approx mv_f, \qquad (6.1)$$

where $v_f$ is the particle's final velocity directed radially inward toward the center line of the LBH trajectory. The energy lost by the LBH equals the energy gained by m

$$\Delta E = \tfrac{1}{2}mv_f^2 = \tfrac{1}{2}m\left(\frac{2GM}{bv_{bh}}\right)^2 = \frac{2G^2M^2m}{b^2v_{bh}^2}, \qquad (6.2)$$

where b is the impact parameter. In this section m represents the target constituents such as the mass of a neutron, a typical air molecule ($N_2$), or a typical rock molecule ($SiO_4$).

We can make a rough estimate of the maximum temperature that the LBH can produce in its wake, assuming that all the energy is converted into heat with negligible heat conduction, and neglecting heats of vaporization and fusion. Thus $\Delta E \approx \tfrac{3}{2}kT_{max}$ in eq. (6.2) implies

$$T_{max} \approx \frac{4G^2M^2m}{3kb^2v_{bh}^2}. \qquad (6.3)$$

Depending on the magnitude of the different variables, it is possible to exceed the melting point of rock ~ 1500 °C. The actual temperature can be much less depending on how much of the energy is partitioned into a shock wave. This would depend on the nature of the part of the earth traversed (e.g. rock, liquid, etc.) and on the magnitude of the energy loss and power input per atom. Thus energy is partitioned differently into heat, tremor, and shock wave.

The LBH energy loss per unit length is

$$\frac{dE}{dx} = N\Delta E\left(\frac{2\pi nb\, db}{N}\right) = \frac{2G^2M^2m}{b^2v_{bh}^2}(2\pi nb\, db) = \frac{4\pi G^2M^2mn}{v_{bh}^2}\int_{b_{min}}^{b_{max}}\frac{db}{b}$$

$$= \frac{4\pi G^2M^2\rho}{v_{bh}^2}\ln\left(\frac{b_{max}}{b_{min}}\right) \qquad (6.4)$$

where N is the number of target particles, $n = N/2\pi b\, db\, dx$ is the number density of the target particles, and $\rho = mn$ is the mass density of these particles.

From eq. (6.4), the total power dissipated by each LBH is thus

$$P = \frac{dE_{total}}{dt} = \frac{NdE}{dx/v_{bh}} = \frac{4\pi G^2 M^2 \rho}{v_{bh}} \ln\left(\frac{b_{max}}{b_{min}}\right). \qquad (6.5)$$

If $\Delta E$ = ionizaton potential $V_i$ of an atom of mass m, substituting $V_i$ into eq. (6.2) yields the ionization parameter

$$b_i = \left[\frac{2G^2 M^2 m}{V_i v_{bh}^2}\right]^{1/2} = \frac{GM}{v_{bh}}\left[\frac{2m}{V_i}\right]^{1/2}. \qquad (6.6)$$

The maximum impact parameter $b_{max} \geq b_i$. Let us next examine a case when the > sign applies.

## 7. TIDAL FORCE IONIZATION

Gravitational and electrostatic (in the case of a charged LBH) tidal force interaction of LBH is an important contributor in the production of polarization and ionization. The tidal force can tear molecules apart and ionize the constituent atoms. Analysis of the effects of the gravitational tidal force on an atomic scale were not available in the scientific literature. A heuristic calculation of tidal force polarization and ionization radii will soon be available (Rabinowitz, 2001 b).

## 8. GRAVITATIONALLY ENHANCED IONIZATION PARAMETER

The intense gravitational field of a LBH causes more atoms to be ionized than given by only kinetic considerations since atoms will be gravitationally captured in orbit around the LBH with the ultimate fate of being ionized even if they do not fall into the BH. This clearly occurs for free particles in the atmosphere, and may also occur if matter is temporarily vaporized along the path of a LBH going through the earth. The gravitational potential energy of a particle of mass m in the field of a LBH is

$$V = -\frac{GMm}{r} - p\left(\frac{GM}{r^2}\right) - \frac{\alpha_p}{2}\left(\frac{GM}{r^2}\right)^2, \qquad (8.1)$$

where p is the permanent dipole moment, and $\alpha_p$ is the gravitational polarizability. This leads (Rabinowitz, 2001 a, b) to an effective ionization radius

$$r_E = b_i \left[1 - \frac{1}{\frac{3}{2}kT}\left(-\frac{GMm}{b_i} - p\left(\frac{GM}{b_i^2}\right) - \frac{\alpha_p}{2}\left(\frac{GM}{b_i^2}\right)^2\right)\right]^{1/2} \qquad (8.2)$$

that is greater than the ionization parameter $b_i$ given by eq. (6.6) so that $b_{max} = r_E \geq b_i$ for a gaseous medium.

If the medium is not gaseous or does not become vaporized, then according to Greenstein and Burns (1984):

$$b_{max} = b_{sonic} = \frac{2GM}{v_{bh}c_s}, \qquad (8.3)$$

where $c_s$ is the speed of sound in the medium.

### 9. MINIMUM IMPACT PARAMETER

The minimum impact parameter is determined by quantum mechanics since quantum effects smear out the particle and reduce the probability of its capture inside the LBH. Though different approaches agree that this occurs when the particle is absorbed by the LBH, they give substantially different values. Fortunately this does not make a big difference in the LBH energy loss per unit length nor in the total power dissipated per LBH as given by eqs. (6.4) and (6.5) since these have a weak logarithmic dependence $\ln(b_{max}/b_{min})$.

From one elementary point of view a target particle cannot be absorbed unless its de Broglie wavelength ≤ the LBH Schwarzchild (horizon) radius

$$b_{min} \sim \lambda = \frac{h}{mv_{bh}} = \frac{2GM}{c^2}, \qquad (9.1)$$

where the relative velocity between the approximately stationary particle and the LBH is the velocity $v_{bh}$ of the LBH. Equation (9.1) would apply provided that $\lambda$ is the shortest length scale in the LBH rest frame.

Another criterion for absorption applies only to very small LBH. It is that $\lambda$ does not change appreciably in a length scale comparable to itself. Interestingly, this implies that $v_{bh} \sim 2GMm/h < c$. Thus for m ~ atomic mass, only LBH with M < $10^{12}$ kg ($R_H = 10^{-15}$ m) can absorb atoms. This criterion would not apply for much larger BH. If this criterion is correct, then even if the classical orbital radius of the particle were small enough to allow it, the particle Compton wavelength for absorption would need to be less than the LBH radius. We can essentially set $b_{min}$ = Compton wavelength:

$$b_{min} \sim \lambda_C = \frac{h}{mc}. \qquad (9.2)$$

A classical orbital approach (Zeldovich and Novikov, 1971) using the Einsteinian effective potential of the LBH which is ~ four times stronger than the Newtonian potential near the LBH yields

$$b_{min} \sim \frac{4GM}{cv_{bh}} \qquad (9.3)$$

for capture and is independent of m as would be expected from the equivalence principle.

## 10. LBH ENERGY LOSS, POWER DISSIPATION, AND RANGE

### 10.1. LBH Orbits Unlikely Inside Earth, Sun, and Neutron Stars

For a closed or quasi-closed (non re-entrant) circular orbit of radius r inside a body of mass density ρ:

$$\frac{Mv_{bh}^2}{r} = \frac{GM[\rho \frac{4}{3}\pi r^3]}{r^2} \Rightarrow r = \frac{v_{bh}}{\left[\frac{4}{3}\pi G\rho\right]^{1/2}}. \qquad (10.1)$$

Equation (6.1) indicates that such orbits execute simple harmonic motion with constant angular velocity for constant ρ, since $\omega = v_{bh}/r = \left[\frac{4}{3}\pi G\rho\right]^{1/2}$ = constant.

For the earth with an average density of ρ = 5.5 x $10^3$ kg/$m^3$, and $v_{bh}$ = 6.2 x $10^5$ m/sec, r = 5.0 x $10^8$ m >> $R_e$ = 6.4 x $10^6$ m. So an internal orbit inside the earth is not possible unless the LBH velocity is greatly reduced.

For the sun with an average density of ρ = 1.4 x $10^3$ kg/$m^3$, and $v_{bh}$ = 8.7 x $10^5$ m/sec, r = 1.2 x $10^9$ m > $R_s$ = 7 x $10^8$ m. So an internal orbit near the limb of the sun would almost be possible.

For a neutron star with an average density of ρ = 4.8 x $10^{17}$ kg/$m^3$, and $v_{bh}$ = 1.6 x $10^8$ m/sec, r = 1.4 x $10^4$ m > $R_n$ = $10^4$ m. So an internal orbit near the limb of the neutron star would almost be possible.

As we shall see in the next sections, velocity degradation of LBH is difficult to achieve by ordinary collisional-like interactions because dE/dx is relatively small. It is also difficult to reduce LBH velocity by particle absorption, since LBH particle absorption is a very low probability event, and when it does occur for particle mass m << M, there is hardly any decrease in $v_{bh}$.

### 10.2. Going Through The Earth

From eqs. (6.6) and (8.2), with an ionization potential ≈ 15 eV = 2.4 x $10^{-18}$ J, and a LBH velocity of 6.2 x $10^5$ m/sec, $10^{12}$ kg LBH have an upper limit $b_{max}= r_E$ ~ 5 x $10^{-4}$ m. If $b_{sonic} \approx b_{ioniz}$, then $b_{max}$ ~ $10^{-8}$ m as given by eqs. (8.3) and (6.6). The minimum impact parameter $b_{min}$ ~ $10^{-17}$m, or $10^{-15}$ m, or $10^{-12}$ m, as given by eqs. (9.2), or (9.1), or (9.3). Because of the logarithmic dependence it does not make much difference which of these $b_{max}$ or $b_{min}$ is used. Thus by eq. (6.4) dE/dx ~ $10^{-2}$ J/m in going through the earth. The overall density of the earth is 5.5 x $10^3$ kg/$m^3$ (5.5 gm/$cm^3$). The mantle density (first 50 miles in from the surface) is 2.7 gm/$cm^3$.

From eq. (6.4), the power dissipated per LBH is

$$P = v_{bh} \frac{dE}{dx} = \frac{4\pi G^2 M^2 \rho}{v_{bh} N} \ln\left(\frac{b_{max}}{b_{min}}\right). \qquad (10.2)$$

Thus P ~ $10^4$ W/LBH for M ~ $10^{12}$ kg. (This is small compared to the total power output of 4.2 x $10^{13}$ W emanating from inside the the earth (Stacey, 1992). From eq. (6.4), the total energy input to the earth per such LBH is

$$E_t \sim \frac{dE}{dx}(\sim R_e) = \left(10^{-2} J/m\right) 6.4 \times 10^6 m \sim 10^5 J/LBH. \quad (10.3)$$

This is insignificant for a LBH with incident velocity of 6.2 x $10^5$ m/sec and kinetic energy of 2 x $10^{23}$ J.

The range of a LBH

$$\Re = \frac{E}{dE/dx} = \frac{\frac{1}{2}M v_{bh}^2}{\left[\frac{4\pi G^2 M^2 \rho}{v_{bh}^2}\right] \ln\left(\frac{b_{max}}{b_{min}}\right)} = \frac{v_{bh}^4}{8\pi G^2 M \rho \ln\left(\frac{b_{max}}{b_{min}}\right)}. (10.4)$$

Equation (10.4) implies that $E = E_o e^{-x/\Re}$. So the range is that path length when the LBH energy has fallen to 1/e of its initial value, $E_o$. The range would be 3 x $10^{25}$ m through solid earth of density 5.5 x $10^3$ kg/$m^3$, which is 21% of the radius of the universe ($R_U$ ~ 1.4 x $10^{26}$ m). To put this into perspective, if the earth had a radius $R_E$ ~ 6 x $10^8$ m (100 times larger than its actual radius), then a LBH with $v_{bh}$ = 6.2 x $10^5$ m/sec in circular orbit just inside this larger earth would make ~ $10^{16}$ revolutions in 1500 billion years i.e. 100 times longer than the present age of the universe. So orbits that are re-entrant into the core of the earth (as well as the sun and neutron stars), could easily persist for almost endless cycles.

### 10.3. Going Through the Sun

Equation (6.4), for an upper limit using the sun's core density of ~ $10^5$ kg/m$^3$ and $\ln(b_{max}/b_{min}) \sim 30$, yields dE/dx ~ $10^{-1}$ J/m for a LBH of M ~ $10^{12}$ kg. From eq. (10.2), the maximum power dissipated in the sun is P ~ $10^5$ W/LBH. Even at this high density the LBH range would be 5 x $10^{25}$ m, which is 38 % of the radius of the universe. If the sun's radius were 1.2 x $10^9$ m (almost a factor of 2 larger than the actual radius $R_s$ = 7 x $10^8$ m), then a LBH with $v_{bh}$ = 8.8 x $10^5$ m/sec in circular orbit just inside this larger sun would make ~ $10^{16}$ revolutions in 2000 billion years i.e. 130 times longer than the present 15 billion-year age of the universe.

### 10.4. Going Through Neutron Stars

The gravitational potential energy of a neutron star is

$$V \sim \frac{GM_n^2}{R_n}, \tag{10.5}$$

where we will take the neutron star mass $M_n$ ~ solar mass = 2 x $10^{30}$ kg, with a radius $R_n$ ~ $10^4$ m. This yields a potential energy of $10^{46}$ J. The binding energy of a neutron of mass $m_n$ = 1.67 x $10^{-27}$ kg is

$$\Delta E_n \sim \frac{GM_n m_n}{R_n} = \frac{GM_n^2}{R_n M_n / m_n} = \frac{10^{46} J}{M_n / m_n} \approx 2.2 \times 10^{-11} \text{ J/neutron.} \tag{10.6}$$

This is $10^2$ MeV which is quite large even compared with nuclear binding energies of 6 - 8 MeV/nucleon. If in eq. (6.6) we set $V_i = \Delta E_n$ ~ $10^{-11}$ J, we obtain $b_{max}$ ~ $10^{-13}$ m. In this view, a LBH displaces a neutron in its gravitational interaction as it goes through a neutron star.

From another point of view, we may think of the interaction of a LBH with a neutron as analogous to the interaction of a LBH in ionizing an atom. A free neutron decays into a proton + electron + antineutrino with a half-life of 10.6 minutes. We may think of the ionization potential of a neutron as $< \sim m_n c^2 - m_p c^2$, the energy difference between the neutron and the proton. Thus $\Delta E_n \sim 939.56$ MeV $- 938.27$ MeV $= 1.29$ MeV. In this scenario, eq.(6.6) yields $b_{max} \sim 10^{-12}$ m.

In the latter scenario and to some degree in the former, $\ln(b_{max}/b_{min}) \sim 10$, and eq. (6.4) yields $dE/dx \sim 10^4$ J/m for a LBH of M $\sim 10^{12}$ kg, and a neutron star density of $5 \times 10^{14}$ kg/m$^3$. The total energy lost is $10^4$ J/m ($\sim 10^4$ m) $\sim 10^8$ J per LBH. The power dissipated is $1.6 \times 10^8$ m/sec ($10^4$ J/m) $\sim 10^{12}$ W.

From eq. (10.4), the range would be $\sim 10^{24}$ m through a neutron star, which is 1% of the radius of the universe. For an orbit just inside a neutron star with $R_n \approx 1.4 \times 10^4$ m, a LBH with $v_{bh} = 1.6 \times 10^8$ m/sec would make $\sim 10^{19}$ revolutions in 0.25 billion year i.e. 1.6 % of the present age of the universe.

## 11. DEVASTATION OF TUNGUS, SIBERIA IN 1908

The devastation of the Tungus region of central Siberia on June 30, 1908, remains a mystery to this day, despite the fact that there were large numbers of eyewitnesses and we know precisely when and where this gigantic explosion took place. A brilliant ball of fire crossed the sky and exploded in the sky with a blast equivalent $\sim 10^{15}$ to $10^{17}$ J ($\sim 30$ million tons of TNT) (Krinov. 1966). More cataclysmic than a hydrogen bomb, the force flattened trees causing them to point radially outward within a 40-mile diameter circle; and hurled creatures like horses to the ground more than 400 miles from Tungus in the area of Kansk.

One of the many speculations that have been considered over the years is that this destruction of an area of more than 1200 square miles was caused by a $\sim 10^{17}$ kg LBH of atomic radius $10^{-10}$ m (Jackson and Ryan, 1973). Burns et al. (1976) conclude that $10^{21}$ to $10^{27}$ J of seismic energy would have been released. This is not only tremendously greater than actually recorded, but greater than some of the largest earthquakes ($\sim 10^{17}$ J) ever recorded. One of the biggest, the 1960 Chilean earthquake, released $> \sim 4 \times 10^{17}$ J, which is large compared with the average annual seismic energy release of $5 \times 10^{17}$ J/yr $= 1.5 \times 10^{10}$ W (Stacey, 1992). Thus Burns et al concluded that the Tungus catastrophe could not have been caused by an LBH. Although Greenstein and Burns (1984) included additional energy release due to Hawking radiation in a later paper (in Fig. 1 in which the $b_{ion}$ scale appears to be low) not related to the Tungus event, this was not included in the papers on Tungus. Also these papers did not consider that LBH might be the missing mass of the universe, this is a distinct possibility in the Rabinowitz model of LBH radiation as discussed in Sec. 4.

Although the conclusion of Barnes et al (1976) may well be correct, one may get much smaller numbers for the total energy released by an LBH going through the earth in Siberia. From eq. (10.3) the energy release for the example $10^{12}$ kg LBH at $6.2 \times 10^5$ m/sec in this paper is only $10^5$ J with a correspondingly larger quadratic effect for larger M LBH. This large disparity results from the scaling of the energy input, where neglecting the logarithmic dependence, $E_t \propto (M/v_{bh})^2$. They assigned the escape velocity $v_{bh} \sim 10^4$ m/sec to their $10^{17}$ kg LBH. Jackson and Ryan (1973) used a similarly low $v_{bh}$ for their $10^{17}$ to $10^{19}$ kg LBH in concluding that, "total energy in the blast wave would be $10^{22}$ to $10^{24}$ erg [$10^{15}$ to $10^{17}$ J]."

As to the large energy release in the atmosphere there are other possibilities besides the impulse energy transfer considered. These include a charged LBH, and (even without Hawking radiation) the explosive disruption of the rotational energy outside an LBH because conservation of angular momentum prevents such matter from falling into the LBH.

So this intriguing question may not have been decided so conclusively as yet. If Jackson and Ryan (1973) are correct, an LBH may still be ingesting part of Siberia since LBH take millions of years to consume objects that are considerably more voluminous than themselves. However it is more likely that if it were an LBH, the LBH went through the earth and exited. Neglecting re-entrant orbits, it is unlikely that such a heavy LBH will come before another $10^5$ years.

## 12. CHANGE IN ANGULAR MOMENTUM DUE TO LBH INTERACTION

### 12.1. General

The total vector sum of the angular momentum of an incident LBH, $L_{bh}$ plus the spin angular momentum S of the target body is conserved because there is no external torque. The final velocity of an LBH as it emerges after travelling a distance r through the target body is

$$v_f = v_{bh}\left[1 - \frac{2\Delta E}{Mv_{bh}^2}\right]^{1/2} \approx v_{bh}\left[1 - 2\frac{2\Delta E}{Mv_{bh}^2}\right] = v_{bh} - \frac{4\Delta E}{Mv_{bh}}, \quad (12.1)$$

where $\Delta E \sim (dE/dx)(\sim r)$, and it was shown above that the second term in the square root factor is $<< 1$. Thus the decrease in the magnitude of the LBH angular momentum is

$$\Delta \vec{L}_{bh} = \vec{d} \times M(\vec{v}_{bh} - \vec{v}_f) \approx dM\frac{4\Delta E}{Mv_{bh}} \sim \frac{4d(rdE/dx)}{v_{bh}}, \quad (12.2)$$

where d is the moment arm with respect to the center of mass of the target body. The initial spin angular momentum of the target body, $S = \tfrac{2}{5}M_t R_t \omega_o^2$, where $\omega_o$ is its initial angular velocity. By conservation of the total angular momentum of the system, $\Delta \vec{S} = -\Delta \vec{L}_{bh}$.

### 12.2. Neutron Star Pulsars

Neutron star pulsars emit pulsed radiation that range in frequencies from x-ray to radio (Davies,1992). The detection of polarization of the radiation, and of the rotation of the plane of polarization within a pulse was an indication that a strong magnetic field plays an important role in the pulses as the neutron star rotates, much like a lighthouse beacon produces a pulse of light in a given direction. The general tendency of pulsars to slow down as well as the cyclotron radiation signature of x-ray pulsars has been explained in terms of huge magnetic fields $\sim 10^6$ to $10^9$ Tesla ($10^{10}$ to $10^{13}$ G). When a star like the sun collapses rapidly with an initial magnetic field of
$10^{-2}$ T, the field gets compressed due to high conductivity followed by a state of extremely high temperature superconductivity, which leads to $B_{final} = B_o\left[r_o^2/r_f^2\right] \sim 10^{-2}$ T $[(10^9 m)^2/(10^4 m)^2] \sim 10^8$ Tesla.

For a $10^{12}$ kg LBH going through a neutron star, with d $\sim$ 0.5 $R_n\sim$ 0.5 x $10^4$ m, r $\sim$ $R_n$, by eq. (12.2) $\Delta S_n$ = $-\Delta L_{bh} \sim 10^4$ kg-m$^2$/sec. This is a relatively small change in angular momentum of the neutron star that could be much larger for a more massive LBH since $\Delta S_n = -\Delta L_{bh} \propto M^2$. Two things may be of interest. One is that this impulse is imparted in a relatively short time $\sim R_n/v_{bh} \sim 10^4$ m/1.6 x $10^8$ m/sec $\sim 10^{-4}$ sec. This is short compared with the 1 to >10 msec period of pulsars. The second is that this can lead to an increase in the pulsar frequency, about half of the time.

Dissipative mechanisms lead to a gradual slowdown of all radio pulsars.  However occasionally there is a sudden increase in frequency, called a starquake, followed by another moderate slowdown of the frequency.  This process is recurrent.  The slowdown time period for most pulsars is $10^3$ to $10^7$ years.   It is thought that the abrupt frequency increase is related to a breakup of the surface crust leading to a decrease in moment of inertia (Shapiro and Teukolsky, 1983).  However it is not clear that such a rearrangement including achievement of a new equilibrium position can occur rapidly enough, and it cannot account for the precipitous frequency increases observed in the Vela pulsar because they occur too frequently (Davies, 1992).

The latter is also a problem for the LBH mechanism unless the LBH are in re-entrant orbits.  In a mechanism that is similar in spirit to this, Stephen Hawking (1971) suggested that a $10^{14}$ kg LBH at the center of a neutron star "would produce a slight shrinking of the surface and might possibly be the cause of the recently observed pulsarquakes." For his mechanism it is also not clear that this process could produce a sufficiently rapid (for most pulsars) and frequent (such as Vela) decrease in the pulsar moment of inertia.  Also both deformation mechanisms should lead to power dissipation of the superconducting currents that maintain the high magnetic field (Rabinowitz, 1976, 1971).

### 12.3. Earth and Sun

Such effects on the earth and on the sun do not appear so clearly because changes in the rotational frequency are not as precisely and dramatically observed.  For the earth a $10^{12}$ kg LBH produces $\Delta S_e = -\Delta L_{bh} \sim 10^5$ kg-m$^2$/sec.  by  eq.(33); and  for the  sun  $\Delta S_s = -\Delta L_{bh} \sim$
$10^{10}$ kg-m$^2$/sec.  Both are small compared with the spin angular momenta of the earth and sun.  With a spin period of 1 day, the earth has $S_e = 7.2 \times 10^{33}$ kg-m$^2$/sec.  For the sun's spin period of 24.7 day, the sun has $S_S = 1.2 \times 10^{42}$ kg-m$^2$/sec.

### 13. Conclusion

The enormous force between BH at close distances as scaled by $R_H$ is surprisingly matched by a comparably large repulsive radiative force.  Whether massive compact objects (herein called LBH) are indeed black holes or Yilmaz grey holes should leave much of the analysis of this paper unchanged.  The Rabinowitz model of LBH radiation in avoiding the unreasonably high radiation of Hawking, permits LBH to be considered as candidates for dark matter in the universe, and ball lightning on earth.  The incidence rate of low mass LBH agrees well with the incidence rate of ball lightning.

In terms of frequency of occurrence, this paper has demonstrated that heavy LBH are unlikely initiators of seismic activity in the earth and in neutron stars unless there is a concentration mechanism such as re-entrant orbits.  In terms of magnitude, it would be possible for very heavy LBH to contribute to seismic activity in the earth and in neutron stars directly or by triggering metastable sites.  LBH are capable of causing abrupt pulsar frequency changes.  A LBH of mass M ~ $10^{12}$ kg was used in these example calculations. Since the energy deposition scales roughly as $M^2$, a heavier LBH can have a correspondingly bigger quadratic effect.  Although it is unlikely that a LBH was responsible for the catastrophic 1908 event in the Tungus region of Siberia, analysis in this paper indicates that past conclusions ruling out a LBH may not be on as firm a basis as formerly thought because important mechanisms were overlooked.

## Acknowledgment

I wish to thank Young Kim for a helpful discussion.